%% file: main.tex
\documentclass{article}
\pdfoutput=1

\usepackage[final]{neurips_2022}

\usepackage[utf8]{inputenc} % allow utf-8 input
\usepackage[T1]{fontenc}    % use 8-bit T1 fonts
\usepackage{hyperref}       % hyperlinks
\usepackage{url}            % simple URL typesetting
\usepackage{booktabs}       % professional-quality tables
\usepackage{amsfonts}       % blackboard math symbols
\usepackage{nicefrac}       % compact symbols for 1/2, etc.
\usepackage{microtype}      % microtypography
\usepackage{xcolor}         % colors
\usepackage{amsmath}
\usepackage{graphicx}
\usepackage{float}
\usepackage{graphicx}

\title{Supervised Pretraining for Molecular \\Force Fields and Properties Prediction}

\author{%
  Xiang Gao, Weihao Gao, Wenzhi Xiao, Zhirui Wang, Chong Wang\thanks{Currently affiliated with Apple Inc., work done at ByteDance Inc.}, Liang Xiang \\
  ByteDance Inc. \\
  \\
  \texttt{\{xianggao,weihao.gao,xiaowenzhi\}@bytedance.com} \\
  \texttt{\{zhirui.wang,xiangliang\}@bytedance.com,mr.chongwang@gmail.com} 
}

\begin{document}

\maketitle

\begin{abstract}
  \input{sections/0.abstract}
\end{abstract}

\input{sections/1.intro}

\input{sections/2.related}

\input{sections/3.method}

\input{sections/4.results}

\input{sections/5.conclusion}

\clearpage
\newpage
{
\small
\bibliographystyle{plainnat}
\bibliography{ref}
}

%\clearpage
%\newpage
%\input{sections/checklist}

%\bibliography{ref}
%\bibliographystyle{icml2021}

%\bibliographystyle{plainnat}
%\bibliography{ref.bbl}

%%%%%%%%%%%%%%%%%%%%%%%%%%%%%%%%%%%%%%%%%%%%%%%%%%%%%%%%%%%%

%\appendix

%\section{Appendix}

%Optionally include extra information (complete proofs, additional experiments and plots) in the appendix. This section will often be part of the supplemental material.

\end{document}

%% file: sections/0.abstract.tex
Machine learning approaches have become popular for molecular modeling tasks, including molecular force fields and properties prediction.
Traditional supervised learning methods suffer from scarcity of labeled data for particular tasks, motivating the use of large-scale dataset for other relevant tasks.
We propose to pretrain neural networks on a dataset of 86 millions of molecules with atom charges and 3D geometries as inputs and molecular energies as labels.
Experiments show that, compared to training from scratch, fine-tuning the pretrained model can significantly improve the performance for seven molecular property prediction tasks and two force field tasks. 
We also demonstrate that the learned representations from the pretrained model contain adequate information about molecular structures, by showing that linear probing of the representations can predict many molecular information including atom types, interatomic distances, class of molecular scaffolds, and existence of molecular fragments. Our results show that supervised pretraining is a promising research direction in molecular modeling%\footnote{The code and pretrained models will be open-sourced on GitHub}.

%. The demonstration covers seven molecular properties datasets and two molecular force field datasets, including a new zero-shot test. New state-of-the-art performance are observed for a few tasks.

% could be helpful.
%Molecular force fields and properties prediction are important tasks for biotechnology, drug discovery, material and energy science. 
%The datasets for these tasks are usually small due to the high cost, and this motivates us to leverage large-scale related data. 

%to learn the relation between two fundamental physical quantities: predicting molecular energies based on their 3D geometries.
%We find that the pretrained model not only performs well in the energy prediction task, but also is helpful for other downstream tasks. 
%Linear probing shows the learned representation can predict the input atom types, interatomic distance, class of molecular scaffolds, and the existence of 85 kinds molecular fragments.

%% file: sections/1.intro.tex
\section{Introduction}

Molecular force fields and properties prediction are important tasks for biotechnology, drug discovery, material and energy science.
Machine learning has become a promising approach for these tasks.
A weakness of supervised machine learning approaches is that they often require a large amount of labeled data to perform well. However, the molecular datasets are usually small due to the high cost to obtain the labels, which may require wet lab experiments or expensive quantum mechanics simulation. 

Pretraining is one approach to alleviate the low-data issue by leveraging relevant large-scale data. The assumption is that the knowledge learned from pretraining can be transferred to the fine-tuning stage on downstream tasks. 
For computer vision, the most popular pretraining strategy is probably a supervised learning approach, where models are pretrained on image classification tasks using large-scale labeled datasets such as ImageNet \cite{russakovsky2015imagenet} and CIFAR \cite{krizhevsky2009cifar}.
For natural language processing, self-supervised pretraining becomes dominant with the success of the masked language modeling \cite{devlin2018bert} and generative pretraining \cite{radford2019gpt2}. 
Recently, pretraining strategies are explored for molecular modeling \cite{hu2019strategies, li2020mpg, rong2020grover, liu2021multiview}, %However, probably due to the lack of large-scale data with proper labels,
and most works employ a self-supervised learning pretraining approach.

Can supervised pretraining help molecular modeling? 
Besides the data availability issue, another challenge is to choose a proper supervised learning target.
Hu et al. \cite{hu2019strategies} experimented with a supervised pretraining strategy, multi-task prediction of more than one thousands biochemical assays. However they observed negative transfer for a few downstream tasks, and suspected that this is due to such supervised pretraining is not "truly-related" to the downstream tasks.
We argue that the supervised pretraining tasks should focus on more fundamental physical properties, instead of the biochemical properties that are measured in a complex environment and are relevant to only a narrow range of downstream tasks. 
We propose to pretrain models to predict molecular energy from molecular structure, with the following considerations.

(i) The supervised pretraining data should be rich. 
Data for fundamental physical properties, such as molecular energies, are often more abundant. For example, the dataset we use in this work, PubChem PM6 dataset \cite{nakata2020pm6}, contains energies calculated for 221 millions molecules.

(ii) The supervised pretraining labels should be accurate rather than noisy. 
Molecular energy can be calculated using established first principle based approach such as density functional theory \cite{parr1983dft} or semi-empirical method \cite{stewart2007pm6method}. Therefore the labels can be obtained by the same mechanism for all samples. In contrast, the experimentally measured biochemical assays used in previous work \cite{hu2019strategies} can be noisy due to the difference in experimental conditions or methods across various data sources.

(iii) The supervised pretraining target should be relevant various downstream tasks.
Many molecular properties of interest are quantities describing the interaction between molecules and the environment, such as proteins in human body or catalyst in batteries. These processes are governed by the forces acting on the atoms. 
Molecular energy is relevant as the force acting on an atom is the negative partial gradients of the molecular potential energy with respect to the coordinate of the atom.
This close relation between molecular energy and molecular structure encourage the pretrained model to understand and represent molecular structures, which is important to perform well on various downstream tasks \cite{nantasenamat2009pqsar}. We empirically find that the pretrained model learns to embed various molecular structural information (see Section~\ref{sec:probe}).

We further propose a force regularization technique, which minimizes the magnitude of the energy gradient with respect to atom coordinates. This makes the pretrained model more suitable for force field prediction tasks. In contrast, instead of using the exact atom coordinates, most previous pretraining tasks \cite{hu2019strategies, li2020mpg, rong2020grover, liu2021multiview} are designed to only use the topological molecular structures. This partially explain why pretraininig approach was not previously used for force field tasks.
\begin{figure*}[h]
  \centering
   \includegraphics[width=0.9\textwidth]{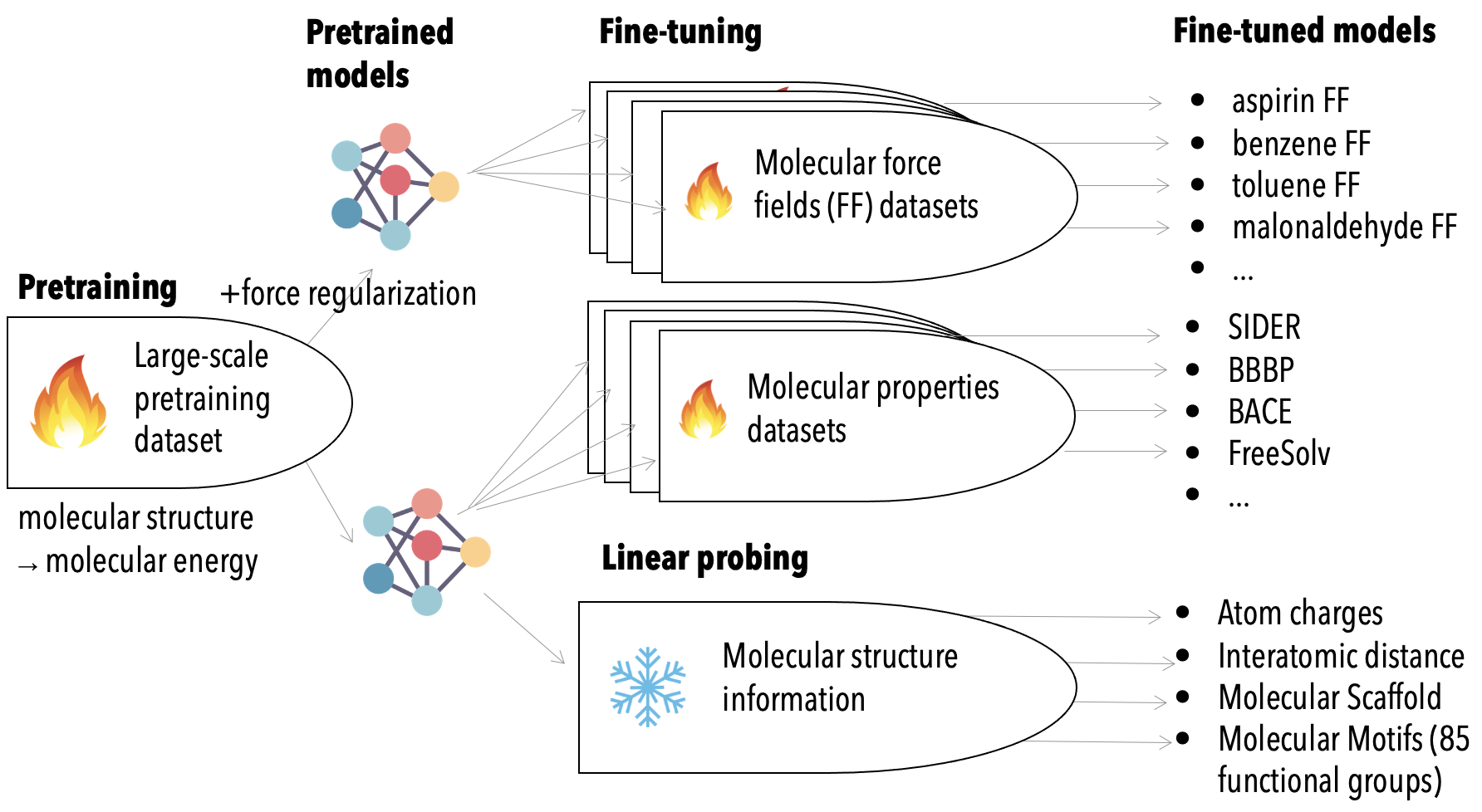}
  \caption{Overview of the proposed pretraining framework.}
  \label{fig:intro}
\end{figure*}

As illustrated in Figure~\ref{fig:intro},
we find that the proposed pretraining strategy can improve the model performance on various downstream tasks, including molecular force fields and molecular properties prediction such as toxicity and molecular water solubility.
%Although the pretrained model is only trained to predict energy, we find it also embed the molecular structure information. 
Linear probing shows the learned representation can predict the input atom types, interatomic distance, molecular scaffolds, and the existence of 85 functional groups.

Our contribution is three-folds. 

1. We propose a supervised pretraining strategy which can improve molecular modeling performance. The pretraining task is to predict molecular energy, a fundamental physical quantity relevant to various downstream tasks and available in large-scale dataset. 

2. We show that, without manual design or constraints, the pretrained model learn to embed the molecular structural information. This is necessary to a wide range of downstream molecular properties prediction tasks. 

3. We show that the proposed pretraining approach helps force fields models generalize better to unseen molecules compared to training from scratch.

%% file: sections/2.related.tex
\section{Background}
% Related Work and Prelimary Concepts

\paragraph{Molecular force field.} This task of predicting the force vector acting on each atom in a molecule. The force depends on the positions of the atoms (conformation). Chmiela et al. \cite{chmiela2017md17} proposed a MD17 dataset containing the energy-conservative force fields data for a few species. An gradient-domain machine learning (GDML) approach is proposed in \cite{chmiela2017md17}. More recently, Klicpera et al.  \cite{klicpera2021gemnet} proposed a GemNet model and obtained better results compared to a few previous works.
Pretraining has not been employed in existing works for this task.

\paragraph{Molecular properties}
There are enormous molecular properties of interest. 
Moleculenet \cite{wu2018moleculenet} collected various datasets of molecular properties on quantum
mechanics, physical chemistry \cite{mobley2014freesolv, delaney2004esol, mendez2019chembl}, biophysics \cite{subramanian2016bace}, and physiology \cite{martins2012bbbp, kuhn2016sider, gayvert2016clintox}. Existing works mostly represent the molecule as graphs, with atoms as node and chemical bonds as edges, and then model the molecules with graph neural networks \cite{kipf2016semi, gilmer2017neural, yang2019analyzing, lu2019molecular} or Transformer\cite{vaswani2017transformer} based architectures \cite{ying2021graphormer, rong2020grover}.

%Recent works \cite{devlin2018bert, radford2019gpt2, russakovsky2015imagenet} employed pretraining to further improve the model performance.

\paragraph{Molecular substructures} Predicting the molecular properties from their structures have been explored extensively \cite{nantasenamat2009pqsar}. Classic methods often investigate molecule structures from two levels, the scaffold \cite{bemis1996scaffold}, and fragments (or motifs), and it is believed that the molecular properties are significantly affected by these substructures \cite{nantasenamat2009pqsar}. The scaffold has been used to group the molecules and split the train/validation/test sets to mimic a more realistic and challenging setting \cite{wu2018moleculenet}. The prediction of the existence of fragments have been used in recent deep learning based methods \cite{rong2020grover} in a un-supervised approach.

\paragraph{Molecular pretraining} 
%has become a popular paradigm in natural language processing (NLP) and computer vision (CV) communities. 
%Models are firstly pretrained on large-scale data and then fine-tuned on downstream tasks usually of much smaller data size. 
%The pretraining can be a self-supervised learning task, such as the masked language modeling task as in BERT \cite{devlin2018bert}, and next-token prediction task in GPT \cite{radford2019gpt2}.
%Another pretraining option is supervised learning, such as the image classification training using large-scale datasets such as ImageNet \cite{russakovsky2015imagenet}.

The pretraining-then-fine-tuning paradigm has been used in molecular modeling, with a focus on self-supervised learning. 
Hu et al. \cite{hu2019strategies} experimented with both node-level and graph-level pre-training. The node-level tasks include prediction of the context of a node, and the prediction of masked node attributes.
For graph-level tasks, they experimented with supervised graph-level property prediction, and graph structural similarity prediction. 
Rong et al. \cite{rong2020grover} demonstrated two self-supervised learning pretraining tasks: contextual property prediction and graph-level motif prediction. 
Li et al. \cite{li2020mpg} proposed a self-supervised pretraining task named pairwise subgraph discrimination, which compares two subgraphs and discriminate whether they come from the same source. 
Liu et al. \cite{liu2021multiview} aligns the latent space learned for molecular graphs and 3D Geometry in attempt to learn knowledge from both input formats.

%The existing works of pretraining for molecular modeling generally focus on self-supervised training. The present work demonstrated that a supervised learning pretraining strategy, predicting molecular energy from molecular structure, can improve various downstream molecular modeling tasks. 

%% file: sections/3.method.tex
\section{Method}

We consider two families of downstream tasks: molecular force field prediction, and molecular properties prediction. 

For the former task, we consider models that take in molecular structure, usually the charges $Z$ (e.g., hydrogen, carbon or oxygen) and 3D coordinates of the atoms $R$, as input, and output the forces $F$ acting on each atom. The force for an atom is a 3D vector, usually sharing the same frame of reference as the atom coordinates.

For the latter task, we consider models that take in molecular structure, and output $y$, which is a scalar value (for regression tasks) or a probability distribution (for classification tasks).

Our approach follows the popular pretraining-and-finetuning paradigm. 

\subsection{Pretraining}

We use the PubChem PM6 dataset \cite{nakata2020pm6} for pretraining. Energy for 86 millions of optimized molecular 3D geometry at neutral states are provided in this dataset. One limitation of our approach is that the performance is affected by the choice of pretraining dataset. The PM6 dataset contain relatively small molecules which makes the pretrained model not suitable for large molecules such as protein or catalysts.

\subsubsection{Molecular force fields}

We train the model to predict the energy $E$  of the optimized 3D molecular geometry $R$. We define a loss term $\mathcal{L}_E$ based on a distance measure, $d$, between the predicted energy $\hat{E}$ and the energy label value $E$. 
\begin{align*}
    \mathcal{L}_E = d \left( E, \hat{E}(Z, R) \right),
\end{align*}
We use mean absolute error as $d$ for $\mathcal{L}_E$.

The PubChem PM6 dataset is built for optimized molecules geometry. The molecular energy is minimized with respect to the atom coordinates. Therefore the energy gradient with respect to the atom coordinates should be close to zero. This motivates us to include a regularization loss term
\begin{align*}
    \mathcal{L}_{\partial \hat{E} / \partial R} = ||  \frac{\partial \hat{E}(Z, R)} {\partial R} ||.
\end{align*}
As this gradient is actually negative potential forces acting on the atoms, we refer $\mathcal{L}_{\partial \hat{E} / \partial R}$ as force regularization.

The pretraining loss for force fields (FF) task is a linear combination of two loss items.
\begin{align*}
    \mathcal{L}_\text{pretrain, FF} = (1 - \alpha) \mathcal{L}_E + \alpha \mathcal{L}_{\partial \hat{E} / \partial R},
\end{align*}
where $\alpha$ is a hyperparameter.

\subsubsection{Molecular properties}

For the molecular properties prediction tasks, many downstream datasets often provide molecular structures in SMILES formats without exact atom coordinates.
We instead use an estimated 3D geometry $R_\text{noisy}$ obtained from SMILES using tools such as RDKit \cite{landrum2013RDKit}.
\begin{align*}
    \mathcal{L}_{E_\text{noisy}} = d \left( E, \hat{E}(Z, R_\text{noisy}) \right),
\end{align*}

In this case the atom coordinates are approximated, so we do not apply the force regularization term. The pretraining loss is
\begin{align*}
    \mathcal{L}_\text{pretrain, properties} = \mathcal{L}_{E_\text{noisy}}.
\end{align*}

\subsection{Fine-tuning}

During fine-tuning, we use the parameters of the pretrained model to initialize the model parameters, except these for the final output layers. For the final output layers parameters, we use random initialization.
%This is because \emph{(i)} the downstream tasks may require outputs of shape different from the pre-training tasks, and  \emph{(ii)} we observe that randomly initializing the final output layers often perform better than using the probably because it provide less biased loss gradient at the beginning of finetuning.

For molecular properties prediction tasks, the training loss measure the distance $d$ between the label output $y$ and predicted output $y$. 
\begin{align*}
  \mathcal{L}_\text{finetune, property} = d (y, \hat{y}),
\end{align*}
where the choice of function $d$ depends on the tasks. We use mean absolute error for regression tasks, and cross entropy loss for classification tasks. 

For molecular force field prediction tasks, we use a linear combination of the energy and force loss.
\begin{align*}
    \mathcal{L}_\text{finetune, FF} = (1 - \gamma) \mathcal{L}_E + \gamma \mathcal{L}_F,
\end{align*}
where $\gamma$ is a hyperparameter.

%% file: sections/4.results.tex
\section{Experiments and discussion}

\input{sections/4-force_field}

\input{sections/4-properties}

\input{sections/4-probing}

%% file: sections/4-force_field.tex
\subsection{Molecular force fields}

For molecular force fields prediction tasks, the existing works \cite{klicpera2021gemnet, chmiela2019sgdml, batzner2021nequip} generally employ an in-domain setting, where the molecule type in training set is the same as the test set. 

Besides this conventional setting, we consider a more challenging, out-of-domain setting. The models are tested on the force field of a unseen molecule. 
This is motivated by distinct characteristics of the pretraining and fine-tuning datasets.
The PubChem PM6 dataset used in pretraining contains a large number of molecules but each only has one conformation. In contrast, the molecular force field dataset contains many conformations for a limited number of molecules.
We wonder if we can combine the knowledge related to different molecules in pretraining dataset and the knowledge of change in conformation in force field dataset. If so, the pretraining should help the model to generalize better on unseen molecules.%, as illustrated in Figure~\ref{fig:unseen}.

We use GemNet-T \cite{klicpera2021gemnet} as the backbone model and use the same set of hyperparameters as \cite{klicpera2021gemnet}. The force is computed as the negative gradient of the predicted energy with respect to atom coordinates. The experiments are conducted with a Nvidia V100 GPU.

\subsubsection{In-domain test}

In this setting, the training data and test data contains different set of conformational geometries and corresponding force fields for the same molecule. We train models for each molecule in the MD17@CCSD dataset \cite{chmiela2018ccsd}. We follow \cite{klicpera2021gemnet} for the train/validation/test split, using 950 samples for training, 50 samples for validation, and 500 samples for test. 
As shown in Table~\ref{tab:ff_in}, the models finetuned from the pretrained model generally perform better than the models trained from scratch.

\input{tables/force_in}

\subsubsection{Out-of-domain test}

In this setting, we employ the MD17 DFT dataset \cite{chmiela2017md17}, which contains 10 kinds of molecules. We construct the training and testing data in a ``train-9-test-1" way. 
%The model is trained on 9 types of molecules and tested on the 1 unseen type of molecule.
For each kind of molecule to be tested, we use the samples from the other 9 molecules as the training set. The test set contains 5k randomly chosen samples, and the training set is built by randomly 50k from each training molecule and mixing the combined 450 samples.

%\begin{figure*}[h]
%  \centering
%   \includegraphics[width=0.8\textwidth]{figures/_unseen.png}
%  \caption{Illustration of the out-of-domain test framework}
%  \label{fig:unseen}
%\end{figure*}

For comparison, we define a na\"ive constant baseline which always predicts zero force. The reported test error for this baseline is equivalent to the element-wise mean absolute value of the force vectors.
As shown in Table~\ref{tab:ff_out}, the test error is generally much higher than the in-domain test. This is expected as the test molecules are not included in the training set. 
Benzene and toluene show a relative low loss, probably because benzene ring, their main structural component, has appeared in several molecules (aspirin, azobenzene, salicylic and naphthalene) in the training set. This implies that the existence of similar substructure in training data significantly help the test performance.

\input{tables/force_out}

The models finetuned from the pretrained model perform significantly better than the models trained from scratch. 
If the pretraining is conducted without force regularization ($\alpha=1$), the improvement become less significant. Force regularization encourage the pretrained model not only to learn the energy for a given geometry, but also the energy gradient at this geometry. This makes pretraining more relevant to the force field prediction tasks.

%% file: tables/force_in.tex
\begin{table}[!ht]
    \centering
    \small
    
    \caption{The mean absolute error (MAE) in meV/Å for MD17@CCSD molecular force fields prediction task in a in-domain setting. }
    \label{tab:ff_in}
    
    \begin{tabular}{p{0.15\textwidth} | p{0.1\textwidth} p{0.1\textwidth} p{0.1\textwidth} p{0.1\textwidth} 
    }
    
    \hline
                   & sGDML \cite{chmiela2019sgdml} & NequIP \cite{batzner2021nequip} & GemNet-T \cite{klicpera2021gemnet} & GemNet-T, finetuned \\
    \hline
    Aspirin        &   33.0  & 14.7   &  10.3  & \textbf{9.3} \\   % one, better than sota 10.3
    Benzene        &    1.7  &  0.8   &  0.8 & \textbf{0.7} \\  % avg, match sota 0.7
    Ethanol        &   15.2  &  9.4   &  \textbf{3.3} & \textbf{3.3} \\ % one, worse than sota 3.1
    Malonaldehyde  &   16.0  & 16.0   &  5.9 & \textbf{5.7} \\   % one, better than sota 5.9
    Toluene        &    9.1  &  4.4   &  2.8 & \textbf{2.6}  \\  % best, better than sota 2.7
    
    \hline
    
\end{tabular}
\label{tab:force_in}
\end{table}

%% file: tables/force_out.tex
\begin{table}[!ht]
    \centering
    \small
    
    \caption{The MAE in meV/Å for MD17@DFT for molecular force fields prediction task in a out-of-domain setting. We train the model on 9 molecules and test on 1 unseen molecule.}
    \label{tab:ff_out}
    
    \begin{tabular} {p{0.25\textwidth} | 
    p{0.1\textwidth} p{0.1\textwidth} p{0.1\textwidth} p{0.1\textwidth} p{0.15\textwidth} }
    
    \hline
    Tested unseen molecule    & Aspirin & Azobenzene & Benzene & Ethanol & Malonaldehyde \\
    \hline
    
    Const. baseline     & 0.899 & 0.910 & 0.626 & 0.841 & 0.907 \\
    From scratch & 0.371 & 0.266 & 0.015 & 0.309 & 0.695 \\
    w/o. force regularization   & 0.248 & 0.220 & 0.013 & 0.318 & \textbf{0.557} \\
    w. force regularization    & \textbf{0.205} & \textbf{0.215} & \textbf{0.013} & \textbf{0.260} & 0.602 \\
    %Rel. Improv. & 44.7\% & 19.1\% & 13.3\% & 15.8\% & 13.4\% \\
    \hline
    
    Tested unseen molecule    & Naphthalene & Paracetamol & Salicylic & Toluene & Uracil \\
    
    \hline
    Const. baseline     & 0.884 & 0.885 & 0.891 & 0.866 & 0.921 \\
    From scratch & 0.118 & 0.174 & 0.231 & 0.092 & 0.278 \\
    From pretrained w/o F  & 0.112 & 0.174 & 0.227 & 0.075 & 0.278 \\
    From pretrained w. F    & \textbf{0.104} & \textbf{0.165} & \textbf{0.191} & \textbf{0.065} & \textbf{0.241} \\
   % Rel. Improv. & 11.9\% & 5.2\% & 17.3\% & 29.3\% & 13.3\% \\
    \hline
    
\end{tabular}
\label{tab:force_out}
\end{table}

%% file: sections/4-properties.tex
\subsection{Molecular properties}

%Similar to previous works \cite{li2020mpg, rong2020grover}, we experiment with multiple small datasets covering molecular physiology, biophysics, and physical chemistry properties.
%The Blood-Brain Barrier Penetration (BBBP) dataset \cite{martins2012bbbp} contains one binary classification task of molecule barrier permeability. The Side Effect Resource (SIDER) dataset \cite{kuhn2016sider} contains 27 binary classification tasks on adverse drug reactions. The ClinTox dataset \cite{gayvert2016clintox} contains two binary classification on drugs failed clinical trials for toxicity reasons. The BACE \cite{subramanian2016bace} dataset contains binding affinity of inhibitors and beta-secretase 1 (BACE-1), an human enzyme. FreeSolv \cite{mobley2014freesolv} lists hydration free energies for small molecules in water. ESOL \cite{delaney2004esol} dataset is used as a regression task to predict molecular water solubility. Lipophilicity (Lipo) dataset, curated from ChEMBL \cite{gaulton2012chembl} database, provides experimental results of octanol/water distribution coefficient. To be consistent with previous works \cite{li2020mpg, rong2020grover}, we use RMSE for regression tasks, and AUC-ROC for classification tasks. All data used in the present work can be downloaded from the MoleculeNet \cite{wu2018moleculenet}. The experiments are conducted with a Nvidia V100 GPU.

Following previous works \cite{li2020mpg, rong2020grover, wu2018moleculenet}, we use scaffold splitting a ratio for train/validation/test as 8:1:1. This splitting method make the molecules in train set do not share molecular scaffold \cite{bemis1996scaffold} with the molecules in validation or test set. The tested molecules are therefore not "similar" to the molecules in training dataset, and this splitting method is believed to offer a more challenging yet realistic way compared to random splitting. For each dataset, we run three times to obtain three different scaffold splits. The average value and standard deviation of the test metrics are reported.

We use EGNN \cite{satorras2021egnn} as the backbone model. If a dataset include more than one tasks (e.g., SIDER contains 27 tasks), we use a multi-task approach, using a single model to predict all tasks simultaneously, similar to \cite{rong2020grover}. The checkpoint cooresponding to the lowest validation loss is evaluated on the test set. We observe that pretraining makes training on downstream task reaches the lowest validation loss much faster than training from scratch, reducing the necessary training epochs from about 300 to 30.

As shown in Table~\ref{tab:molnet}, the models finetuned from the pretrained model generally perform much better than the models trained from scratch. 
We do not observe negative transfer as Hu et.al. \cite{hu2019strategies} did with their multi-task supervised pretraining strategies. 
Compared to self-supervised pretraining methods (N-GRAM \cite{liu2019n}, Hu et.al. \cite{hu2019strategies}, GROVER \cite{rong2020grover}, and MPG \cite{li2020mpg}), our supervised pretraining method achieve similar or better results. 
%This suggests that the pretrained model learned to embed the information necessary for these downstream tasks. 

%\subsubsection{Ablation of the pretraining dataset}
%\input{tables/ablation}

\input{tables/moleculenet}

%% file: tables/moleculenet.tex
\begin{table}[h]

    \centering
    % https://tex.stackexchange.com/questions/56008/different-sizes-of-font-available-in-table
    \small
    %\footnotesize
    %\scriptsize
    \caption{The performance comparison for molecular properties prediction tasks. The numbers in brackets are the standard deviation.}
    \label{tab:molnet}
    
% \resizebox{0.9\textwidth}{!}
{
\renewcommand{\arraystretch}{1.1}
\tabcolsep=0.06cm
\begin{tabular}{cccccccc}
\hline
& \multicolumn{4}{|c}{Classification (AUC-ROC)} & \multicolumn{3}{|c}{Regression (RMSE)}\\
\hline
Dataset & \multicolumn{1}{|c}{BBBP} & {SIDER} & {ClinTox} & {BACE} & \multicolumn{1}{|c}{FreeSolv} & {ESOL} & {Lipo} \\
\# Molecules & \multicolumn{1}{|c}{2039} & {1427} & {1478} & {1513} & \multicolumn{1}{|c}{642} & {1128} & {4200}  \\
%\# Task & 1 & 27 & 2 & 1 \\
\hline
TF\_Robust \cite{ramsundar2015massively} & $0.860_{(0.087)}$ & $0.607_{(0.033)}$ & $0.765_{(0.085)}$ & $0.824_{(0.022)}$ & $4.122 _{(0.085)}$ & $1.722 _{(0.038)}$ & $0.909 _{(0.060)}$ \\
GraphConv \cite{kipf2016semi}& $0.877_{(0.036)}$ & $0.593_{(0.035)}$ & $0.845_{(0.051)}$ & $0.854_{(0.011)}$ & $2.900_{(0.135)}$ & $1.068_{(0.050)}$ & $0.712_{(0.049)}$ \\
Weave  \cite{kearnes2016molecular}& $0.837_{(0.065)}$ & $0.543_{(0.034)}$ & $0.823_{(0.023)}$ & $0.791_{(0.008)}$ & $2.398_{(0.250)}$ & $1.158_{(0.055)}$ & $0.813_{(0.042)}$ \\
SchNet \cite{schutt2017schnet}& $0.847_{(0.024)}$ & $0.545_{(0.038)}$ & $0.717_{(0.042)}$ & $0.750_{(0.033)}$ & $3.215_{(0.755)}$ & $1.045_{(0.064)}$ & $0.909_{(0.098)}$ \\
MPNN  \cite{gilmer2017neural}& $0.913_{(0.041)}$ & $0.595_{(0.030)}$ & $0.879_{(0.054)}$ & $0.815_{(0.044)}$ & $2.185_{(0.952)}$ & $1.167_{(0.430)}$ & $0.672_{(0.051)}$ \\
DMPNN \cite{yang2019analyzing}&  $0.919_{(0.030)}$ &  $0.632_{(0.023)}$ & $0.897_{(0.040)}$ & $0.852_{(0.053)}$ & $2.177_{(0.914)}$ & $0.980_{(0.258)}$ & $0.653_{(0.046)}$ \\
MGCN  \cite{lu2019molecular}& $0.850_{(0.064)}$ & $0.552_{(0.018)}$ & $0.634_{(0.042)}$ & $0.734_{(0.030)}$ & $3.349_{(0.097)}$ & $1.266_{(0.147)}$ & $1.113_{(0.041)}$ \\
AttentiveFP \cite{xiong2019pushing}& $0.908_{(0.050)}$ & $0.605_{(0.060)}$ &  $0.933_{(0.020)}$ &  $0.863_{(0.015)}$ & $2.030_{(0.420)}$ &  $0.853_{(0.060)}$ &  $0.650_{(0.030)}$ \\
\hline
N-GRAM \cite{liu2019n}& $0.912_{(0.013)}$ & $0.632_{(0.005)}$ & $0.855_{(0.037)}$ &  $0.876_{(0.035)}$ & $2.512_{(0.190)}$ & $1.100_{(0.160)}$ & $0.876_{(0.033)}$ \\
Hu. et.al\cite{hu2019strategies} & $0.915_{(0.040)}$ & $0.614_{(0.006)}$ & $0.762_{(0.058)}$ & $0.851_{(0.027)}$ & - & - & - \\
GROVER\cite{rong2020grover} & 
\textbf{0.940}$_{(0.019)}$ & 
${0.658}_{(0.023)}$ & 
${0.944}_{(0.021)}$ & 
${0.894}_{(0.028)}$ & $1.544_{(0.397)}$ & $0.831_{(0.120)}$ &  {0.560}$_{(0.035)}$ \\
MPG \cite{li2020mpg} & 
$0.922_{(0.012)}$ & 
\textbf{0.661}$_{(0.007)}$ & 
\textbf{0.963}$_{(0.028)}$ &
\textbf{0.920}$_{(0.013)}$ & 
$1.269_{(0.192)}$ &
 $0.802_{(0.023)}$ & 
 ${0.576}_{(0.029)}$ \\
\hline
EGNN \cite{satorras2021egnn} & 
0.896$_{(0.030)}$ &     % bbbp, ft3 
0.646$_{(0.015)}$ &     % sider, all
0.779$_{(0.089)}$ &     % clintox, ft3
0.868$_{(0.046)}$ &     % bace, ft3
1.421$_{(0.239)}$ &     % freesolv r3
0.651$_{(0.032)}$ &     % esol, r3
0.738$_{(0.072)}$  \\   % lipo, avg
EGNN, finetuned & 
\textbf{0.948}$_{(0.010)}$ &    % bbbp, t3 
\textbf{0.665}$_{(0.011)}$ &    % sider, t3 
\textbf{0.974}$_{(0.017)}$ &    % clintox, t3 
\textbf{0.934}$_{(0.016)}$ &    % bace, t3
\textbf{0.844}$_{(0.025)}$ &    % freesolv r3
\textbf{0.511}$_{(0.013)}$ &    % esol, r3
\textbf{0.510}$_{(0.017)}$  \\  % lipo, avg
\hline
\end{tabular}
}
\vspace{-2ex}
\label{tab:properties}
\end{table}

%% file: sections/4-probing.tex
\subsection{What have the pretrained models learned?}
\label{sec:probe}

The experiments above show that pretraining on energy prediction tasks improve the model performance on both force fields and molecular properties prediction.
The improvement on the force fields seems not surprising as the energy is closely related to force, making the pretraining task relevant to the downstream tasks.
However, the relation between energy and the tested molecular properties is not so obvious. This makes us wondering what knowledge the pretrained models have learned.
Both the molecular energy and the molecular properties depend on the molecular structure \cite{nantasenamat2009pqsar, katritzky2010quantitative}. This motivates us to test whether the pretrained model learns to embed the structural information, which can be then used in downstream tasks for molecular properties prediction. 

Following \cite{alain2016probe}, we employ linear probes to analyze representation learned by the pretrained EGNN model.
Each layer in EGNN model output a node representation $h \in \mathbb{R}^D $ for each atom, where $D$ is the hidden dimension. The node representation can be summed up as the graph representation. 
For linear probing, the pretrained model is frozen. We use a trainable linear layer to map the learned representation to the output space. 

We consider the following two categories of diagnostic tasks to test whether the pretrained model learns to embed the molecular structural information.

\subsubsection{Reconstructing the input}

\begin{figure}[h]
  \centering
   \includegraphics[width=0.8\textwidth]{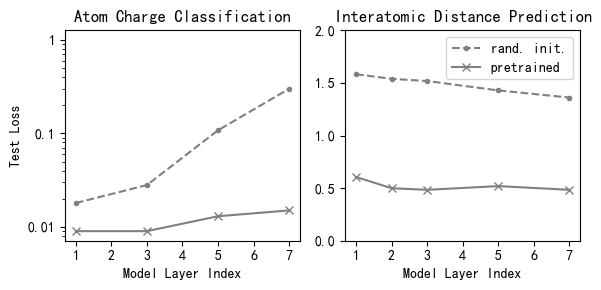}
  \caption{Linear probing on input reconstruction tasks}
  \label{fig:recon}
\end{figure}

We firstly test whether the representation produced by the pretrained model can be used to reconstruct the input of EGNN, the atom types and interatomic distance.

\paragraph{Atom charge classification.} We start with a simple test, whether the node representation contains the information about atom charges (e.g. hydrogen, carbon or oxygen). For a molecule with $n$ atoms, the node representation for the $i$-th atom, $h_i$, is sent to a linear probe layer to output $p_i$ the probability distribution of atom charges. 
%The linear probe layer is parameterized by $W_e \in \mathbb{R}^{n_e \times D}$, where $n_e=118$ is the number of considered atom charges. This linear probe layer is trained with negative log likelihood loss $\mathcal{L}_\text{elem}$. 
%\begin{align*}
%    p_i &= \text{Softmax}( h_i W^T_e ), \\
%    \mathcal{L}_\text{elem} &= - \frac{1}{n} \sum^n_{i=1} \log p_{i,e_i}.
%\end{align*}

\paragraph{Interatomic distance prediction.} This task tests whether the pretrained model learns to embed the 3D geometries. $h_i$ is input to a linear probe layer to output $\hat{x}$, the estimated 3D coordinates. 
%The linear probe layer is parameterized by $W_x \in \mathbb{R}^{3 \times D}$
%The pairwise interatomic distance $d_{ij}$ is computed and the normalized error is used as the training loss $ \mathcal{L}_\text{dist}$.
%\begin{align*}
%    \hat{x_i} &= h_i W^T_x, \\
%    \hat{d}_{ij} &= || \hat{x}_i - \hat{x}_j || , 
%    d_{ij} = || x_i - x_j ||, \\
%    \mathcal{L}_\text{dist} &= \sqrt{ \frac{1}{n(n-1)} \sum^n_{i=1} \sum^n_{j=1, j\neq i} \frac{(d_{ij} - \hat{d}_{ij})^2} {d^2_{ij}} }.
%\end{align*}

We compare the test error of these three diagnostic tasks with EGNN models of random parameters.
As illustrated in Figure~\ref{fig:recon}, the probing performance of the pretrained model is significantly better than model of random parameters. This indicates that the pretrained model learns to embed structure information, although it is only trained to predict molecular energy.

For atom type classification task, the test loss using the representation from the first layer is smaller than that of the last layer, as shown in Figure~\ref{fig:recon}. This is expected as in EGNN, the atom type information is only directly provided to the first layer. Model of random parameters quickly lose this information as indicated by the significantly higher loss. In contrast, the pretrain model learned to pass this information to the last layer with smaller loss.
For interatomic distance prediction task, as the atom coordinates information is sent to each EGNN layer, the test loss do not change significantly across different layers. 

The results above show that, although only trained to predict molecular energy, the pretrained model learn to embed molecular structural information in the learned representation. As structural information is important to various downstream tasks, pretraining improve the performance compared to training from scratch.

\subsubsection{Identifying substructure}

\begin{figure}[h]
  \centering
   \includegraphics[width=0.8\textwidth]{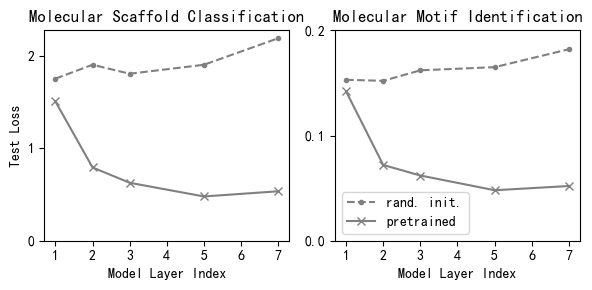}
  \caption{Linear probing on molecular substructure prediction tasks}
  \label{fig:sub}
\end{figure}

We then test if the pretrained model learns to embed the information necessary to identify the molecular substructures. 

\paragraph{Molecular scaffold classification.} We are interested in higher level molecular structure, scaffold. From the training data, we choose the top 100 frequent scaffold. With trainable linear layers, parametrized by $W^T_sh \in \mathbb{R}^{D \times D}$ and $W^T_sp \in \mathbb{R}^{100 \times D}$, The learned representation is used to predict the scaffold class probability distribution $p_s$. Given labels $y_\text{scaffold}$, the probe layers are trained with negative log likelihodd loss.
%\begin{align*}
%    h_\text{scaffold} &= \frac{1}{n} \sum^n_{i=1} (h_i W^T_sh), \\
%    p_s &= \text{Softmax}( h_\text{graph} W^T_sp ), \\
%    \mathcal{L}_\text{scaffold} &= - \log p_s(y_\text{scaffold}).
%\end{align*}

\paragraph{Molecular motif identification.} Similar to \cite{rong2020grover}, we consider 85 functional groups \footnote{Full list: http://rdkit.org/docs/source/rdkit.Chem.Fragments.html}, such as aliphatic carboxylic acids and H-pyrrole nitrogens, as the molecular motifs. We test if the pretrained model embed the information to identify whether each of these fragment appear in the molecule. We formulate this as a multi-task binary classification problem.
With trainable linear layers, parametrized by $W^T_fh \in \mathbb{R}^{D \times D}$, $W^T_\text{negative} \in \mathbb{R}^{85 \times D}$, and $W^T_\text{positive} \in \mathbb{R}^{85 \times D}$, the learned representation is used to predict the binary classification probability $p_j$ for the $j$-th fragment. Given labels $y_\text{motif}$, the probe layers are trained with negative log likelihodd loss.
%\begin{align*}
%    h_\text{motif} &= \frac{1}{n} \sum^n_{i=1} (h_i W^T_fh), \\
%    h_\text{negative} &= h_\text{motif} W^T_\text{negative}, \\
%    h_\text{positive} &= h_\text{motif} W^T_\text{positive} \\
%    p_j, &= \text{Softmax}(h_{\text{negative},j}, h_{\text{positive},j}), \\
%    \mathcal{L}_\text{motif} &= - \frac{1}{85} \sum^{85}_{j=1} \log p_j(y_{\text{motif},j}).
%\end{align*}

Similar to the input reconstruction tasks, the probing performance of the pretrained model is significantly better than model of random parameters, as shown in Figure~\ref{fig:recon}.
For molecular scaffold classification task, using the representation from the last EGNN layer of the pretrained model results in a much smaller test error compared to that of the first layer, as shown in Figure~\ref{fig:recon}. Scaffold class is not a input of the EGNN model, and have to be predicted based on the molecular global structure. This implies that the pretrained model learns to gradually obtain the global representation of the molecules at later layers. 
Similar trends are observed for molecular fragment identification task. As the information pass along to deeper layers, the representation embed more global information and is easier for prediction of molecular subtructure tasks. 
%Rong et al. \cite{rong2020grover} proposed to use these 85 molecular motifs as pretraining labels. However, we show that, although only trained to predict molecular energy, our pretrained model is still able to embed the information necessary to complish these tasks designed in previous self-supervised training works.

%% file: sections/5.conclusion.tex
\section{Conclusion}

We proposed and demonstrated a supervised learning pretraining strategy for molecular force fields and properties prediction. 
We find that the pretrained model not only learns to perform the energy prediction task, but also embed the molecular structure information. 
Experiments show that, compared to training from scratch, fine-tuning the pretrained model can significantly improve the performance. The demonstration covers seven molecular properties datasets and two molecular force field datasets, including a new zero-shot test. New state-of-the-art performance are observed for a few tasks.